# Rotation of galaxies and dark matter

V. A. Golovko

Moscow Polytechnic University

Bolshaya Semenovskaya 38, Moscow 107023, Russia

E-mail: fizika.mgvmi@mail.ru

Abstract

In a previous paper by the author was proposed a new metric for the gravitational field of a thin rotating disk physically different from the Kerr metric. The metric is admissible for any angular momentum of the disk. As demonstrated in the present paper the parameter determining the angular momentum of the Milky Way greatly exceeds its gravitational radius so that the Kerr metric physically admissible only if the angular momentum is sufficiently small is completely inapplicable to the Milky Way. It is shown on the basis of the new metric that the rotation of the Milky Way plays a decisive role in the motion of satellites in its gravitational field. The effects due to the rotation can imitate the presence of hypothetical dark matter.

*Keywords*: General relativity; Rotation of galaxies; Milky Way; Dark matter.

# 1. Introduction

In Ref. [1] (hereafter referred to as I) were proposed metrics for the gravitational field of a charged and rotating mass which are physically different from the Kerr-Newman one. The metric the most adequate in this case is given in Eq. (3.1) of I. The metric is singular only at the surface of an infinitely thin rotating disk which creates the gravitational field. And what is more important, the metric is admissible for any angular momentum of the disk in contradistinction to the Kerr-Newman metric which is physically admissible only if the angular momentum is sufficiently small. Therefore the new metric opens up possibilities for studies of the rotation of galaxies, in which case the Kerr-Newman metric is completely inapplicable because of a large value of the angular momentum of the galaxies as shown in Appendix A with the Milky Way. Seeing that celestial bodies are practically noncharged, we shall restrict our consideration to the situation where the central rotating disk and other bodies are not charged. [1]

In this paper we investigate the rotation of the Milky Way Galaxy and its influence on the Milky Way's gravitational field. The Milky Way will be approximated by an infinitely thin rotating disk. Although this approximation is rather simplified and does not reflect the structure of the Milky Way's disk, the approximation enables one to find out principal effects due to the rotation. Unexpectedly the effects are clearly pronounced and imitate the presence of hypothetical dark matter. In addition the effects explain the existence of the central bulge of the Milky Way.

In Sec. 2 we write out equations of I required for our studies. Section 3 is devoted to the treatment of the radial motion in the gravitational field of the Milky Way, and Sec 4 is concerned with the motion in its equatorial plane. Remarks as to the results obtained are made in the concluding section.

# 2. Basic equations

In the present paper we employ the metric given in Eq. (3.1) of I, the metric describing the gravitational field created by an infinitely thin rotating disk. As mentioned in Introduction we consider noncharged bodies. This being so, we put $r_q = 0$ in Eqs. (3.2)–(3.3) of I. Now Eqs. (3.1)–(3.3) of I acquire the form

---

[1] If the rotating disk is charged, its charge should not be too large. The restriction on the charge is not relevant to general relativity but is due to the limits of applicability of classical electrodynamics used in this case [1].

$$ds^2 = \frac{\Lambda}{\Sigma}c^2dt^2 - \frac{r_g^4 \, e^{-2r_g/r} \, \Sigma}{r^4\left(1-e^{-r_g/r}\right)^4 \Delta}dr^2 - \Sigma d\theta^2 - \frac{Y}{\Sigma}\sin^2\theta \, d\varphi^2 + \frac{2aZ}{\Sigma}\sin^2\theta \, cdtd\varphi, \quad (2.1)$$

where

$$\Delta = \frac{r_g^2 \, e^{-r_g/r}}{\left(1-e^{-r_g/r}\right)^2} + a^2, \quad \Sigma = \frac{r_g^2}{\left(1-e^{-r_g/r}\right)^2} + a^2\cos^2\theta, \quad \Lambda = \frac{r_g^2 \, e^{-r_g/r}}{\left(1-e^{-r_g/r}\right)^2} + a^2\cos^2\theta,$$

$$Y = \left[\frac{r_g^2}{\left(1-e^{-r_g/r}\right)^2} + a^2\right]^2 - \Delta a^2 \sin^2\theta, \quad Z = \frac{r_g^2}{1-e^{-r_g/r}}. \quad (2.2)$$

We recall that $r_g = 2Gm/c^2$ is the gravitational radius of the disk of mass $m$ and of radius $a$, and the rotation of the disk is characterized by the angular momentum $L = amc$. The metric is singular only at $r = 0$, the value $r = 0$ corresponding to the surface of the rotating disk.

For the present studies we need the equations of motion for the metric. Since we imply noncharged bodies, we put $r_q = \varepsilon = 0$ in Eqs. (3.11)–(3.14) of I with the result

$$c\frac{dt}{ds} = \frac{\beta Y - ahZ}{\Delta\Sigma}, \quad (2.3)$$

$$\left(\frac{dr}{ds}\right)^2 = \frac{r^4 \, e^{2r_g/r}}{r_g^4 \Sigma^2}\left\{\beta\left[r_g^2 + a^2\left(1-e^{-r_g/r}\right)^2\right] - ah\left(1-e^{-r_g/r}\right)^2\right\}^2$$

$$- \frac{r^4 \, e^{2r_g/r}\left(1-e^{-r_g/r}\right)^2 \Delta}{r_g^4 \Sigma^2}\left[r_g^2 + \kappa\left(1-e^{-r_g/r}\right)^2\right], \quad (2.4)$$

$$\left(\frac{d\theta}{ds}\right)^2 = \frac{1}{\Sigma^2}\left[\kappa - a^2\cos^2\theta - \left(\beta a\sin\theta - \frac{h}{\sin\theta}\right)^2\right], \quad (2.5)$$

$$\frac{d\varphi}{ds} = \frac{1}{\Delta\Sigma}\left(\frac{h\Lambda}{\sin^2\theta} + a\beta Z\right). \quad (2.6)$$

In these equations, $\beta = E_0/(\mu c^2)$ where $E_0$ is the energy of a test particle and $\mu$ is its mass, $h = L_0/(\mu c)$ where $L_0$ is the angular momentum of the test particle. In other words, the constants $\beta$ and $h$ are the dimensionless energy (the total energy divided by the rest energy of the test particle) and a magnitude of the angular momentum of the particle (with dimension of length), respectively. The last constant $\kappa$ (with dimension of square of length) is related to the constant $K$ of Landau and Lifshitz [2], problem 1 in § 104, by $\kappa = K/(\mu^2 c^2)$.

We also write down the expression for the nontensorial velocity $v$ of the test particle measured by an observer stationed at infinity and given in Eq. (3.17) of I:

$$v^2 = c^2 \frac{\Delta^2 \Sigma^2 (\beta^2 \Sigma - \Lambda)}{\Lambda(\beta Y - ahZ)^2}. \quad (2.7)$$

## 3. Radial motion

The radial motion in the above metric is considered in I. Here we analyze the motion in more detail. If $a \neq 0$, the radial motion of a test particle is possible only along the axis of rotation of the gravitating disk where $\theta = 0$. From (2.5) we see that now $h = 0$ and $\kappa = a^2$. Substituting this into (2.3) and (2.4) results in the following equation for the dependence of the coordinate $r$ upon the time $t$

$$\left(\frac{dr}{dt}\right)^2 = c^2 \frac{r^4 e^{2r_g/r}\left(1-e^{-r_g/r}\right)^4 R_\Delta^2}{\beta^2 r_g^4 R_\Sigma^2} R_1, \tag{3.1}$$

in which

$$R_\Delta = r_g^2 e^{-r_g/r} + a^2\left(1-e^{-r_g/r}\right)^2, \quad R_\Sigma = r_g^2 + a^2\left(1-e^{-r_g/r}\right)^2,$$

$$R_1 = \beta^2 - \frac{r_g^2 e^{-r_g/r} + a^2\left(1-e^{-r_g/r}\right)^2}{r_g^2 + a^2\left(1-e^{-r_g/r}\right)^2}. \tag{3.2}$$

In Eq. (3.1), only the factor $R_1$ can be negative. Recalling that $\beta$ is the dimensionless energy of a test particle, in parallel with Landau and Lifshitz [2], problem 1 in § 102, we introduce a positive dimensionless "effective potential energy" (the positive effective potential energy divided by the rest energy of the test particle)

$$U(r) = \left[\frac{r_g^2 e^{-r_g/r} + a^2\left(1-e^{-r_g/r}\right)^2}{r_g^2 + a^2\left(1-e^{-r_g/r}\right)^2}\right]^{1/2}, \tag{3.3}$$

so that $R_1 = \beta^2 - U^2(r)$. The motion of the test particle is possible only if $\beta \geq U(r)$ when $R_1 \geq 0$.

To examine $U(r)$ we calculate the derivative

$$\frac{dU}{dr} = \frac{r_g^3 e^{-r_g/r}}{2r^2 U} \frac{r_g^2 - a^2\left(1-e^{-r_g/r}\right)^2}{\left[r_g^2 + a^2\left(1-e^{-r_g/r}\right)^2\right]^2}. \tag{3.4}$$

It follows from this ($r \neq 0$) that, if $a < r_g$, then always $dU/dr > 0$. If $a > r_g$, the effective potential energy $U(r)$ is a minimum at

$$r = -\frac{r_g}{\ln\left(1-\dfrac{r_g}{a}\right)}. \tag{3.5}$$

If $a \gg r_g$, the minimum is at $r \approx a$, which amounts to saying that in this case the position of the minimum is practically independent of the gravitational constant $G$.

Figure 1 shows the behavior of $U(r)$ for different values of the parameter $a$ that determines the angular momentum of the gravitating disk. If $a > r_g$, we have a potential well (this conforms with the results of I). The particle oscillates in the well along the axis of rotation of the disk. The well comes into being only if the rotation of the disk is sufficiently rapid.

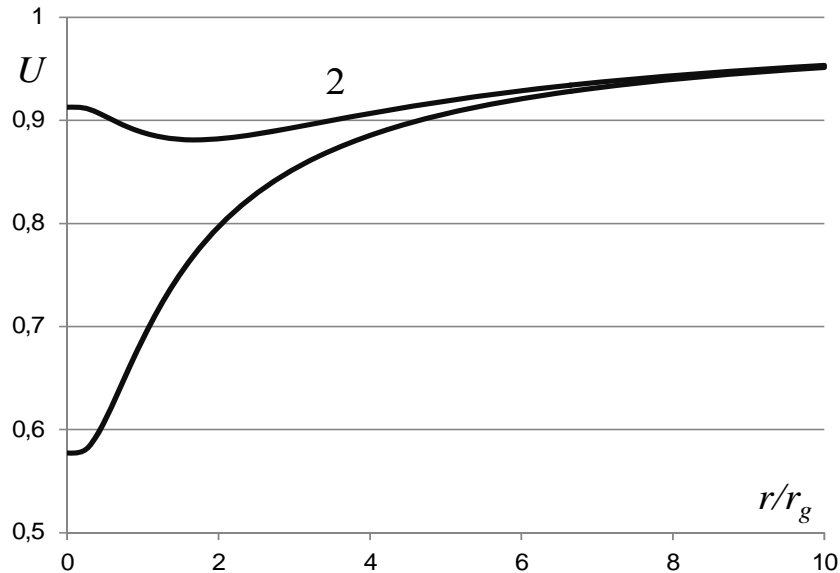

**Fig. 1**. Dependence of the effective potential $U(r)$ of (3.3) upon $r$ in the radial motion perpendicularly to the disk. Curve 1: $a/r_g = 0.5$, curve 2: $a/r_g = 5$.

As to the Milky Way, Appendix A shows that $a > r_g$ for it and thereby there is a potential well on the axis of rotation. Neighboring stars are captured by the well, which gives rise to formation of a bulge. Consequently the central bulge of the Milky Way is due to the Milky Way's rotation. To obtain a comprehensive picture of the potential well and not only on the axis of rotation it is necessary to analyze the equations of motion when $0 \leq \theta < \pi/2$, which is beyond the scope of the present paper. Here we can only remark that the motion along the axis of rotation is unstable because an arbitrary small perturbation perpendicular to the axis will lead the star away from the axis. The stable orbits in the central bulge of the Milky Way are parallel to its disk, which is considered at the end of Sec. 4.

The central bulge cannot be explained by the presence of dark matter near the axis of the Milky Way's rotation, dark matter attracting neighboring stars, if the above potential well is absent. In the absence of such a potential well, dark matter attracted by the disk of the Galaxy will necessarily fall upon the disk. Therefore the central bulge of the Milky Way and of other rotating galaxies is completely due to their rapid rotation when the above potential well is formed. To explain the formation of this potential well the metric of (2.1) should be employed, which, on the other hand, confirms the adequacy of the metric for description of the gravitational field of rotating bodies. It should be underlined that the stars in the central bulge cannot rotate around the centre of the Galaxy because in this case they must cross the disk of the Galaxy where they will be inevitably absorbed by the disk heavily populated near the Galaxy's centre. The stable orbits of the stars in the central bulge should not touch the disk, which is possible only if the above potential well exists.

## 4. Motion in the equatorial plane

We turn now to the motion in the equatorial plane where $\theta = \pi/2$. First of all we write out the formula for the velocity $v$ of a star rotating about the Milky Way, the velocity measured by an observer at infinity. If we put $\theta = \pi/2$ and substitute the quantities of (2.2) into (2.7), we have

$$v^2 = c^2 \frac{\left[r_g^2 e^{-r_g/r} + a^2\left(1 - e^{-r_g/r}\right)^2\right]^2 (\beta^2 - e^{-r_g/r})}{\left[\beta r_g^2 + \beta a^2(2 - e^{-r_g/r})\left(1 - e^{-r_g/r}\right)^2 - ah\left(1 - e^{-r_g/r}\right)^3\right]^2} e^{r_g/r}. \quad (4.1)$$

Let us compare this with the case where the gravitating mass does not rotate being a point mass, of course. We mention in passing that this case is considered in Ref. [3]. If $a = 0$, Eq. (4.1) yields

$$v^2 = c^2 \frac{(\beta^2 - e^{-r_g/r})}{\beta^2} e^{-r_g/r}. \quad (4.2)$$

We see a drastic difference between (4.1) and (4.2) when $r \to 0$. According to Eq. (4.1) $v \to \infty$ as $r \to 0$ whereas according to Eq. (4.2) $v \to 0$ in the same limit (recall that the value $r = 0$ corresponds to the surface of the rotating disk). It is clear from the physical point of view that the rotating and gravitating disk must drag neighboring objects into rotation. Although the Milky Way is not, of course, a solid disk and the speeds of orbiting stars cannot be infinite; nevertheless the results following from (4.1) demonstrate that the speed of the orbiting stars near the Milky Way can be rather high in contradiction with the Newtonian law of gravitation. The contradiction found experimentally is commonly ascribed to the presence of dark matter. We see that the contradiction may be due to the rotation of the Milky Way.

We are coming now to the study of the motion in the equatorial plane. Once $\theta = \pi/2$, we see from (2.5) that $\kappa = (\beta a - h)^2$. Substituting this into (2.4) yields

$$\left(\frac{dr}{ds}\right)^2 = \frac{r^4 e^{2r_g/r}}{r_g^4 \Sigma^2} R_\pi, \tag{4.3}$$

where

$$R_\pi = \left[\beta(r_g^2 + a^2 y^2) - ahy^2\right]^2 - \left[r_g^2(1-y) + a^2 y^2\right]\left[r_g^2 + (\beta a - h)^2 y^2\right]. \tag{4.4}$$

To simplify the notation we have introduced the quantity

$$y = 1 - e^{-r_g/r}, \quad 0 \le y \le 1. \tag{4.5}$$

If necessary, the derivative $dr/dt$ can be found from (4.3) with use made of (2.3). It is worthy of remark that, in the present case and in other cases considered in the paper, Eq. (2.3) leads to no peculiarities.

One can introduce an "effective potential energy" $U(r)$ as above only if an expression of the type $R_\pi$ of (4.4) does not contain the first power of $\beta$. If one considers the motion of a star in the gravitational field of the Milky Way, account must be taken of the fact that the angular momentum of the star characterized by the parameter $h$ is very small as compared to the angular momentum of the Milky Way given by the parameter $a$ so that we can neglect $h$ in (4.4) ($h \ll \beta a$) with the result:

$$R_\pi = r_g^2 \left[r_g^2 + a^2\left(1 - e^{-r_g/r}\right)^2 \left(2 - e^{-r_g/r}\right)\right]\left[\beta^2 - U^2(r)\right], \tag{4.6}$$

where

$$U(r) = \left[\frac{e^{-r_g/r} + \xi\left(1 - e^{-r_g/r}\right)^2}{1 + \xi\left(1 - e^{-r_g/r}\right)^2 \left(2 - e^{-r_g/r}\right)}\right]^{1/2}. \tag{4.7}$$

Here we have returned to the previous notation and introduced the positive parameter

$$\xi = \frac{a^2}{r_g^2}. \tag{4.8}$$

For use later we recast Eq. (4.1) in terms of $y$ and $\xi$:

$$v^2 = c^2 \frac{(1 - y + \xi y^2)^2 (\beta^2 - 1 + y)}{(1-y)\left[\beta + \beta\xi(1+y)y^2 - ahy^3/r_g^2\right]^2}. \tag{4.9}$$

Before analyzing Eq. (4.7) it is instructive to write Eq. (2.6) in the present case where $h \ll \beta a$ (also $h\Lambda \ll \beta aZ$ as $\Lambda = Z$ when $r \gg r_g$ and $\theta = \pi/2$):

$$\frac{d\varphi}{ds} = \frac{a\beta Z}{\Delta\Sigma}. \tag{4.10}$$

As long as $d\varphi/ds$ is positive (we assume throughout the paper that $a > 0$), the test particle rotates in the same sense as the gravitating disk and the speed of rotation is proportional to $a$, which amounts to saying that the speed of rotation is very high. Classical mechanics is completely inapplicable to this motion because the particle has a nonzero angular velocity ($d\varphi/ds \neq 0$) whereas its angular momentum is nil ($h = 0$).

In order to investigate the potential energy $U(r)$ of (4.7) we rewrite Eq. (4.7) in terms of $y$ and $\xi$ as

$$U(y) = \left[\frac{1 - y + \xi y^2}{1 + \xi y^2 (1 + y)}\right]^{1/2}. \tag{4.11}$$

Differentiating we have

$$\frac{dU}{dy} = -\frac{\xi^2 y^4 + 2\xi y^2 (1 - y) + 1}{2U\left[1 + \xi y^2 (1 + y)\right]^2}. \tag{4.12}$$

Seeing that $0 \leq y \leq 1$, this derivative is always negative whereas $dU/dr = (dU/dy)(dy/dr)$ is nonnegative because $dy/dr \leq 0$ ($dy/dr = -r_g e^{-r_g/r}/r^2$; $dy/dr = 0$ if $r = 0$). As a result, $U(r)$ increases monotonically from $[\xi/(1+2\xi)]^{1/2}$ when $r = 0$ to 1 as $r \to \infty$. The qualitative behavior of $U(r)$ as a function of $r$ can be schematically represented by a curve similar to curve 1 in Fig. 1. Therefore circular orbits are impossible since $U(r)$ has no extrema if $r \neq 0$.

If $\beta \geq 1$, the test particle spirals from infinity into the rotating disk because the angle $\varphi$ augments monotonically in view of (4.10). If $\beta \geq 1$, the parameter $\beta$ is related to the velocity of the test particle at infinity $v_\infty$ by Eq. (3.13) of [3]:

$$\beta = \frac{1}{\sqrt{1 - v_\infty^2/c^2}}. \tag{4.13}$$

This relation follows from (4.1) as well.

When $\beta < 1$, the test particle spirals into the rotating disk from the apogalaction, that is, from the most remote point in the orbit where $U(r) = \beta$. As an example, we discuss the situation where $\beta \approx 1$. In this situation the motion near the apogalaction should be described by classical mechanics because $v \ll c$ there. The equation $U(r) = \beta$ can be written as

$$\beta^2 y^3 - (1 - \beta^2) y^2 + \frac{y}{\xi} - \frac{1 - \beta^2}{\xi} = 0. \tag{4.14}$$

We have here two small parameters $1 - \beta^2$ and $1/\xi$ (the parameter $\xi$ is considered in Appendix A). The solution to (4.14) in this case is found in Appendix B and is

$$y = (1 - \beta^2)\left[1 + \xi(1 - \beta^2)^3\right]. \tag{4.15}$$

Seeing that $y \approx 0$, we put $y = 0$ in (4.9) where it is possible with the result

$$v^2 = c^2 \frac{\beta^2 - 1 + y}{\beta^2}. \tag{4.16}$$

We substitute (4.15) with account taken of the fact that $y \to r_g/r$ as $r \to \infty$ by (4.5):

$$v^2 = c^2 \frac{\xi(1-\beta^2)^4}{\beta^2} \approx c^2 \xi y^4 = c^2 \frac{a^2 r_g^2}{r^4}. \tag{4.17}$$

This formula is not consistent with classical mechanics where $v^2 \propto 1/r$. At the same time, $v^2$ of (4.17) may be rather large because of $c^2 a^2$. Of course, when $r \to 0$, the velocity $v$ tends to infinity inasmuch as $y \to 1$ in (4.9). As a result, we see that the Milky Way's rotation affects the motion of its satellites even at great distances owing to $a$ in (4.17).

Another case where it is possible to introduce an effective potential energy as above is furnished by the relation $\beta a - h = 0$. The relation $h = \beta a$ signifies that the angular momentum of the object in question is comparable to the one of the Milky Way provided that $\beta$ is not too small. Therefore the case in point is a galaxy orbiting the Milky Way. It should be remarked that the mass of the galaxy may be rather small because the value of $h$ is proportional to the distance between the Milky Way and the galaxy which can be rather great. So, this case matches to a dwarf galaxy located at a specific average distance rotating about the Milky Way. In this case Eq. (4.9) becomes

$$v^2 = c^2 \frac{(1 - y + \xi y^2)^2 (\beta^2 - 1 + y)}{(1-y)(1+\xi y^2)^2 \beta^2}. \tag{4.18}$$

Once $\beta a - h = 0$, Eq. (4.4) can be written in the form

$$R_\pi = r_g^4 \left[\beta^2 - U^2(r)\right], \tag{4.19}$$

where

$$U(r) = \left[e^{-r_g/r} + \xi\left(1 - e^{-r_g/r}\right)^2\right]^{1/2}. \tag{4.20}$$

Equation (2.6) yields now

$$\frac{d\varphi}{ds} = \frac{a\beta(\Lambda + Z)}{\Delta\Sigma}. \tag{4.21}$$

Here again $d\varphi/ds > 0$, as in (4.10), and the speed of rotation is proportional to $a$.

The qualitative behavior of the effective potential energy $U(r)$ of (4.20) as a function of $r$ is represented in Fig. 2. It follows from the figure that, if $\beta > \sqrt{\xi}$, the satellite galaxy spirals from infinity into the Milky Way's disk. When $\sqrt{\xi} > \beta > 1$, the galaxy spiraling from infinity does not reach the disk going around it and spirals to infinity again. If $1 > \beta > \sqrt{1 - 1/4\xi}$, the galaxy when starting from the apogalaction spirals about the disk and returns to the apogalaction. Finally, if $\beta = \sqrt{1-1/(4\xi)}$, the dwarf galaxy rotates about the disk in circular orbit with radius $r_0 = -r_g/\ln[1-1/(2\xi)]$.

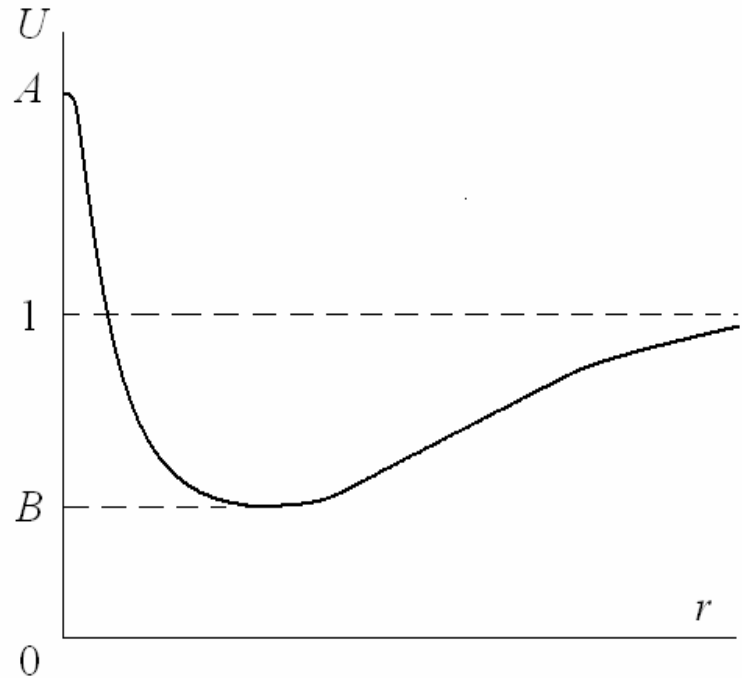

**Fig. 2**. Schematic dependence of the effective potential $U(r)$ of (4.20) upon $r$ (not to scale). Point A corresponds to $\sqrt{\xi}$, point B corresponds to $\sqrt{1-1/4\xi}$. The curve has $\xi > 1$.

The equation that determines the position of the apogalaction and perigalaction is $\beta^2 = U^2(r)$, which is seen from (4.19) and (4.3) (at these positions $dr/ds = 0$). We write the equation as

$$\xi y^2 - y + 1 - \beta^2 = 0. \qquad (4.22)$$

The solution to the equation is

$$y = \frac{1 \pm \sqrt{1 + 4\xi(\beta^2 - 1)}}{2\xi}. \qquad (4.23)$$

Recalling that $y \geq 0$ we see that, when $\beta > 1$, the equation has only one positive solution; and when $\beta < 1$, it has two positive solutions existing if $\beta \geq \sqrt{1-1/4\xi}$. This is consistent with Fig. 2.

As above we shall assume that $\beta \approx 1$. Moreover, we suppose that $|\beta^2-1|$ is so small that $\xi|\beta^2-1| \ll 1$. In this situation the fist approximate solution valid for $\beta > 1$ and $\beta < 1$ is

$$y_1 = \frac{1}{\xi}. \qquad (4.24)$$

This solution corresponds to the perigalaction. When $\beta < 1$, the second approximate solution is

$$y_2 = (1-\beta^2)[1 + \xi(1-\beta^2)]. \qquad (4.25)$$

This solution corresponds to the apogalaction which exists when $\beta < 1$.

Considering the motion in the vicinity of the perigalaction we should take into account the fact that $y_1$ of (4.24) is very small [$1/\xi \sim 10^{-4}$ according to (A.5)]. Consequently we can set $y = 0$ in (4.18) where it is possible. In the remaining expression ($\beta^2 - 1 + y$) we use (4.22) so that

$$v^2 = c^2 \frac{\xi y^2}{\beta^2}. \tag{4.26}$$

Substituting $y_1$ of (4.24) with $\beta \approx 1$ yields

$$v = \frac{c}{\sqrt{\xi}}. \tag{4.27}$$

If we put $\xi \sim 10^4$ here, we shall obtain the speed of the satellite galaxy at the perigalaction equal to $v \sim 10^{-2}c = 3000$ km/s. This enormous speed is due to our approximation according to which the Milky Way is considered to be a solid disk. At the same time this result demonstrates that in the real situation the speed of the satellite galaxy may be rather high.

Considering the motion in the vicinity of the apogalaction we remark that $y_2$ of (4.25) is also very small and thereby Eq. (4.26) holds. Instead of substituting this $y_2$ which essentially depends on $\beta$, we refer to (4.5) and see that $y$ is small when $r \to \infty$, and $y \to r_g/r$ in this case. We place this last $y$ with $\beta \approx 1$ in (4.26) to obtain

$$v^2 = c^2 \frac{a^2}{r^2}. \tag{4.28}$$

It is interesting to note that the Newtonian gravitational consonant $G$ disappears from this formula and the role of $G$ goes over to $a^2$. As to the dependence of $v^2$ on $r$ we can make the same comment as the one concerning Eq. (4.17).

We turn now to the circular orbit which occurs when $\beta = \sqrt{1 - 1/4\xi}$. It follows from (4.23) that then $y = 1/(2\xi)$. Introducing these $y$ and $\beta$ into (4.18) yields

$$v^2 = c^2 \frac{2\xi(4\xi - 1)}{(2\xi - 1)(4\xi + 1)^2}. \tag{4.29}$$

If we take $\xi \sim 10^4$ as in (A.5), we shall have $v \sim 1500$ km/s. This speed is of the same order of magnitude as the speed resulting from (4.27).

There is yet another case where it is possible to introduce an effective potential energy $U(r)$ as above. We see from (2.5) that $\theta = $ constant if

$$\kappa = a^2\cos^2\theta - \left(\beta a\sin\theta - \frac{h}{\sin\theta}\right)^2. \tag{4.30}$$

We shall also suppose that

$$h = \beta a \sin^2\theta. \tag{4.31}$$

Now (4.30) yields $\kappa = a^2 \cos^2\theta$. If $\theta$ is small, one has $h \ll a$. Therefore the case in point will be a star orbiting the Milky Way.

Placing these $\kappa$ and $h$ in (2.4) we see that we can introduce $U(r)$ which is

$$U(r) = \left[ \frac{e^{-r_g/r} + \xi\left(1 - e^{-r_g/r}\right)^2}{1 + \xi \cos^2\theta \left(1 - e^{-r_g/r}\right)^2} \right]^{1/2}. \qquad (4.32)$$

If $\theta = 0$, we have (3.3); if $\theta = \pi/2$, we have (4.20). Consequently, when $\theta$ changes from 0 to $\pi/2$, we go from Fig. 1 to Fig. 2. In both cases the potential energy $U(r)$ has a minimum and therefore circular orbits are possible. The orbits lie in planes parallel to the equatorial plane. When $\theta$ is small, we shall have a star moving in the central bulge of the Milky Way.

## 5. Concluding remarks

The present paper shows that the rotation of the Milky Way plays a decisive role in the motion of stars or other galaxies in its gravitational field. For example, it is just the rotation that explains the existence of the central bulge of the Milky Way. The effect of the rotation may be so strong that the quantity $a^2$ characterizing the rotation takes the place of the gravitational constant as is seen from Eq. (4.28). Of course, the results obtained in the paper can be applied to other rotating galaxies. The effects due to the rotation can imitate the presence of hypothetical dark matter. It may be that the rotation of the galaxies and of clusters of galaxies influences the evolution of the Universe as long as there must exist an additional energy due to the rotation. Besides, the Universe or its big parts may rotate as a whole and centrifugal forces arising in this case can accelerate their expansion imitating the presence of dark energy.

The study of the rotation of the galaxies has become possible when the new metric for the gravitational field of a rotating mass of Ref. [1] was found. The metric is admissible for any angular momentum of the rotating mass whereas the angular momentum of the galaxies can be very large as shown in Appendix A with the Milky Way. The well-known Kerr metric is completely inapplicable in this situation.

In this paper we employ rather a simplified model for the Milky Way representing it as an infinitely thin rotating disk. When considering the motion of satellites in the vicinity of the Milky Way the model can yield only qualitative results. At great distances from the Milky Way its structure is not of crucial importance so that the model may produce quantitative results. As is seen from Appendix A the radius $a$ of the disk modeling the Milky Way is small in comparison with its radius $R$. It would be unreasonable to put $a = R$ because this would yield an enormous angular momentum $L = Rmc$ and an enormous angular velocity $\omega = L/I$ for the Milky Way, which would essentially overestimate all effects due to its rotation.

We have, in the paper, considered several examples where solutions to relevant equations are rather simple. Nevertheless even these simple examples demonstrate that without regard for the rotation it is impossible to comprehend the actual gravitational field of the Milky Way.

Of course, it would be desirable to confirm the results of the paper by comparison with observational data. At this stage of theoretical investigations, however, it will be useless because it is clear in advance that there will be no agreement between the results obtained in the paper and the observational data. This is due to the fact that the Milky Way is not an infinitely thin rotating disk where the velocity of a test particle tends to infinity at the disk's surface. The main aim in the paper is to attract attention to the fact that the rotation of galaxies plays an important role in their gravitational field and to stimulate further investigations along this line. It is necessary to develop a more realistic model of the rotating Milky Way in order to compare theoretical results and observational data.

The results of the present paper do not deny the existence of dark matter. However, properties and the distribution of dark matter cannot be correctly studied if the rotation of galaxies is not taken into proper account, which follows from the results of the paper.

**Appendix A. Parameters for the Milky Way**

The mass of the Milky Way can be estimated as $m = 1.29 \cdot 10^{12}\ m_\odot$ [4]. Therefore the gravitational radius of the Milky Way is

$$r_g = \frac{2Gm}{c^2} = 3.8 \cdot 10^{12}\,\text{km}. \qquad (A.1)$$

The angular momentum of the Milky Way taken from Ref. [5] is $L = 0.97 \cdot 10^{67}$ J·s. Karachentsev [5] uses $m = 1.5 \cdot 10^{11}\ m_\odot$ and thereby the actual angular momentum of the Milky Way may be greater than the above value. With use made of $L = 0.97 \cdot 10^{67}$ J·s we obtain for the parameter $a$ of the Milky Way that

$$a = \frac{L}{mc} = 1.3 \cdot 10^{13}\,\text{km}. \qquad (A.2)$$

It should be remarked that in order to be consistent with Karachentsev's calculations one ought to utilize his mass $m = 1.5 \cdot 10^{11}\ m_\odot$ with the result

$$a = 1.1 \cdot 10^{14}\,\text{km}. \qquad (A.3)$$

We see that in any case $a > r_g$ and even $a \gg r_g$ for the Milky Way. If the gravitational field possesses its own angular momentum, this will only augment the value of $a$.

There is another way to evaluate the angular momentum of the Milky Way. We can compute the moment of inertia $I$ with the help of the formula $I = mR^2/2$ for a uniform disk and obtain $L$ from $L = I\omega$. There are different estimates for the radius $R$ of the Milky Way. We take $R = 4.7 \cdot 10^{17}$ km although some authors propose a larger value. The angular velocity $\omega$ of the outer parts of the Milky Way is close to the one in the neighborhood of the Sun. The latter is $\omega = 8.94 \cdot 10^{-16}$ s$^{-1}$ [6]. These numbers give

$$a = 3.3 \cdot 10^{14} \text{ km}. \tag{A.4}$$

This estimate is close to (A.3). With this $a$, we calculate the parameter $\xi$ of (4.8):

$$\xi = \frac{a^2}{r_g^2} \sim 10^4. \tag{A.5}$$

When considering the motion of Milky Way's satellites in the equatorial plane it is preferable to imply (A.4) or (A.5) because it is the rotation of the outer parts of the Milky Way with the angular velocity $\omega$ which plays a leading part in the gravitational field created by the Milky Way in this plane.

It may be noted that the Kerr metric is physically admissible only if $a \leq r_g/2$. Consequently the Kerr metric is absolutely unfit for the gravitational field of the Milky Way.

**Appendix B. Equation (4.14)**

Equation (4.14) can be recast as

$$-(1-\beta^2)y^3 + [y - (1-\beta^2)]y^2 + \frac{y - (1-\beta^2)}{\xi} = 0. \tag{B.1}$$

Instead of $y$ we introduce $\bar{y}$ according to $y = 1-\beta^2 + \bar{y}$ and obtain the equation

$$-(1-\beta^2)(1-\beta^2+\bar{y})^3 + \left[(1-\beta^2+\bar{y})^2 + \frac{1}{\xi}\right]\bar{y} = 0. \tag{B.2}$$

The first term here is small since we assume the smallness of $1-\beta^2$. Upon neglecting this term we arrive at a first approximation $\bar{y} = 0$. We rewrite Eq. (B.2) in the form

$$\bar{y} = \frac{(1-\beta^2)(1-\beta^2+\bar{y})^3}{(1-\beta^2+\bar{y})^2 + 1/\xi}. \tag{B.3}$$

Placing the first approximation $\bar{y} = 0$ in the right-hand side we obtain the second approximation

$$\bar{y} = \frac{(1-\beta^2)^4}{(1-\beta^2)^2 + 1/\xi}. \tag{B.4}$$

If we return to $y$, we shall have

$$y = 1 - \beta^2 + \frac{\left(1-\beta^2\right)^4}{\left(1-\beta^2\right)^2 + 1/\xi}. \tag{B.5}$$

The quantity $1-\beta^2$ can be arbitrarily small whereas $\xi$ is fixed. Therefore we can neglect $1-\beta^2$ in the denominator, which leads to Eq. (4.15).